\documentclass[preprint2]{aastex}
\slugcomment{submitted to {\it The Astronomical Journal}}
\begin{document}
\title{$uvbyCa$H$\beta$ CCD Photometry of Clusters. IV. Solving the Riddle
of NGC 3680}
\author{Barbara J. Anthony-Twarog\altaffilmark{1} and 
Bruce A. Twarog\altaffilmark{1}}
\affil{Department of Physics and Astronomy, University of Kansas, 
Lawrence, KS 66045-7582}
\affil{Electronic mail: bjat@ku.edu,twarog@ku.edu}
\altaffiltext{1}{Visiting Astronomer, Cerro Tololo Interamerican Observatory.
CTIO is operated by AURA, Inc.\ under contract to the National Science
Foundation.}
\begin{abstract}
CCD photometry on the intermediate-band $uvbyCa$H$\beta$ system is presented
for the open cluster, NGC 3680. Restricting the data to probable cluster members
using the CMD and the photometric indices alone defines a sample of 34 stars
at the cluster turnoff that imply $E(b-y)$ = 0.042 $\pm$0.002 (s.e.m.) or $E(B-V)$
= 0.058 $\pm$0.003 (s.e.m.), where the errors refer to internal errors alone.
With this reddening, [Fe/H] is derived from  both $m_1$ and $hk$ using both
$b-y$ and H$\beta$ as the temperature indices. The agreement among the four
approaches is excellent, leading to final value of [Fe/H] = --0.14 $\pm$0.03 for the
cluster and removing the apparent discrepancy between the past $uvby$ analyses
and extensive results from the red giants. The primary source of the photometric
anomaly appears to be a zero-point offset in the original $m_1$ indices. 
Using the homogenized and combined $V$, $b-y$ data from a variety of studies
transformed to $B-V$, the cluster CMD is compared to NGC 752, IC 4651, 
and the core-convective-overshoot
isochrones of \citet{GI02}. By interpolation to the proper metallicity, it is found that
the $E(B-V)$, (m-M), and age for NGC 752, IC 4651, and NGC 3680 are
(0.03, 8.30, 1.55 Gyr), (0.10, 10.20, 1.7 Gyr), and (0.06, 10.20, 1.85 Gyr),
respectively. The revised age and metallicity sequence and the color distribution 
of the giants provide evidence for the suggestion that the giants defining the apparent
clump in NGC 3680 are predominantly first-ascent giants, as indicated by their
Li abundance, while the clump stars in NGC 752, 0.1 mag bluer in $(B-V)$,
are He-core-burning stars. When combined with the color distribution in IC 4651,
it is suggested that over this modest age range where He-core flash becomes
important, the distribution of  so-called clump stars switches from being dominated by
He-core burning stars to first-ascent giants in the bump phase.
\end{abstract}

\section{INTRODUCTION}
This is the fourth paper in an extended series detailing the derivation
of fundamental parameters in star clusters using precise intermediate-band
photometry to identify probable cluster members and to calculate the
cluster's reddening, metallicity, distance and age. The initial motivation
for this study was provided by \citet{TAT97}, who used a homogeneous 
open cluster sample to identify structure within the galactic abundance 
gradient, structure that has been corroborated most recently through 
the use of Cepheids by \citet{AN12,AN02,LU03}, though the origin 
and reason for the survival of the feature remains elusive \citep{SC01,MI02,LP03}. 

Detailed justifications of the program and the observational approach adopted
have been given in previous papers in the series \citep{AT00a,AT00b,TW03}
(hereinafter referred to as Papers I, II, and III)
and will not be repeated. Suffice it to say that the reality of the galactic 
features under discussion will remain questionable unless the error 
bars on the data are reduced to a level smaller than the size of the effect 
we are evaluating and/or the size of the sample is statistically enhanced. 
The overall goal of this project is to do both.

The focus of this paper is the intermediate-age open cluster, NGC 3680. 
By the normal standards of open cluster research, NGC 3680 has 
been well-studied on a variety of photometric systems, including 
$BV$ \citep{EG69,AT91,KO97}, DDO \citep{MC72,CL83}, Washington \citep{GE91}, 
and $uvby$H$\beta$ \citep{NI88,ATS89,NO96,BR99}. It has a respectable 
level of membership information via proper motions \citep{KO95} and 
radial velocities \citep{ME95,NO97}. It has also been
analyzed using moderate-dispersion spectroscopy \citep{FJ93,FR02} 
and at high dispersion \citep{PA01}. It was initially included in the program 
as a source of standard stars for calibration of the CCD intermediate-band photometry.
However, in the study of \citet{TAT97}, it was exceptional in that the 
abundance estimates from DDO photometry and moderate-dispersion 
spectroscopy of the giants both disagreed significantly with 
the $uvby$-based abundance from stars at the turnoff of the cluster. 
The giants indicated a cluster with [Fe/H] near  --0.15 while the turnoff 
stars produced [Fe/H] closer to +0.1. Given the
large data samples and the small internal errors in the estimates, the
difference could not be dismissed as a byproduct of the internal errors.
This left three possible solutions:

(a) The $uvby$ system is inherently flawed and, for some unknown reason,
produces cluster parameters that are distortions of reality. Though one cannot
rule this out without an independent means of testing the parameters 
generated by the $uvby$ system for clusters, the extensive applications of
the system to field stars over the last 35 years have generated no evidence
for such a failure beyond the usual revisions in the calibrations as
data on all sides have improved, contrary to the claims of some authors
\citep{PA01}. In particular, for F stars of disk metallicity, the type found 
at the turnoff of NGC 3680, the parametric calibrations have been repeatedly 
tested and revised \citep{CR75,SN89,ED93} because of the interest in 
applying the techniques to studies of the chemical and dynamical 
evolution of the disk. 

(b) The cluster giants and dwarfs produce different results because the
stars are different; the distribution of elements in the evolved giants
has been altered by evolution while the main sequence stars remain
pristine samples of the initial cluster abundance. This could be a
plausible suggestion in the case of DDO photometry where the metallicity
index includes a CN band, but fails to explain the spectroscopic results
tied to, among other things, the Fe lines. Moreover, one is faced with the
prospect of explaining why this difference appears in no other open cluster
for which comparable data are available, in agreement with standard
post-main-sequence evolution scenarios for stars of intermediate mass.

(c) The simplest option, given the high internal precision of the data, is
that a zero-point error exists within the sample, either in the $uvby$H$\beta$
photometry or in the DDO data. The latter case seems less likely since
the DDO [Fe/H] was derived independent of the spectroscopic data and both
agree at a level consistent with what is found for other clusters. To
test the possibility of a zero-point problem with the $uvby$ data, it
was decided to reduce and analyze the cluster as a program object, rather
than include it within the calibration of the CCD data.

Section 2 contains new photoelectric observations of stars in the
field of NGC 3680 on the $Caby$ system, the details of the 
$uvbyCa$H$\beta$ CCD observations, and their
reduction and transformation to the standard system. In
Sec. 3 we discuss the CMD and begin the process of identifying the
sample of probable cluster members. Sec. 4 contains the derivation of the
fundamental cluster parameters of reddening and metallicity. In
Sec. 5, these are combined with broad-band data to derive the distance and
age through comparisons with other clusters and with theoretical isochrones
while testing the post-main-sequence predictions of the models. 
Sec. 6 summarizes the status of NGC 3680 in the context of 
constraining current models of stellar evolution.

\section{The Data}

\subsection{Observations:Photoelectric $Caby$}
The $Caby$ system was initially defined and developed using traditional
photoelectric photometry obtained at a number of observatories, but
predominantly CTIO and KPNO, between
1983 and 1996. These observations, primarily of field stars, have been
discussed and analyzed in a series of papers
\citep{ATT91,TAT95, ATT00} culminating most recently
in the definition of the system for Hyades main sequence stars \citep{AT02}.
In addition to the Hyades, a number of stars were observed in open
and globular clusters with the intent of providing internal
standards for future CCD work. Included in this sample were 8 stars in the
field of NGC 3680, observed with a pulse-counting system equipped with an
S-20 photomultiplier on the 1.0 m and 1.5 m telescopes at CTIO 
between 1989 and 1992. The cluster
stars were transformed and reduced with the field stars and should be well
tied to the standard system. Details on the reduction and merger of the
photometry over a series of nights and runs may be found in \citet{TAT95}
and will not be repeated here. Suffice it to say that the $V$ and $(b-y)$
values are on the system of \citet{OLS93}. A summary of the data for 8 stars
is given in Table 1 where the errors refer to the standard error of 
the mean for each index. The stars will not be
used in the calibration of the CCD photometry but will serve, instead, as
a check on the reliability of the overall procedure for calibrating the
$Caby$ CCD system.

\subsection{Observations: CCD $uvbyCa$H$\beta$}
The new photometric data for NGC 3680 were obtained using the Cassegrain-focus 
CCD imager on the National Optical Astronomy Observatory's 0.9-m telescope
at Cerro Tololo Interamerican Observatory. We used a Tektronix 2048 by 2048 
detector at the $f/13.5$ focus of the telescope, with CTIO's $4'' \times 4''$ $uvby$ filters 
and our own $3'' \times 3''$ H$\beta$ and $Ca$ filters.  The field size is 
$13.5'$ on a side.  Frames of all seven filters were obtained in both 
May 2000 and Jan. 2002.

\subsection{Reduction and Transformation}
Paper I includes a fairly comprehensive description of the steps used
to process the data and the procedures by which we merge photometry based
on profile-fitting routines to produce average magnitudes and indices of very 
high internal precision. Presented here is an outline of the steps followed to 
transform these instrumental indices to standard photometric systems.  
We observe standard stars selected from the photometry catalogs 
of \citet{OLS83,OLS93,OLS94} and our own catalog of $Caby$ indices, \citet{TAT95};
we also observe stars in open clusters that may be used as standards
each photometric night.  The clusters are ideal for determining
extinction coefficients for indices other than H$\beta$. Aperture magnitudes
are obtained for every standard star in the field or in a cluster, as well
as for uncrowded stars in the program cluster fields.  Apertures ranged from
8 to 11 pixels centered on the star, surrounded by sky annuli of
comparable enclosed area.  Following correction for atmospheric extinction
and the exclusion of stars with serious crowding, average
magnitudes in each bandpass are constructed for all stars.
From these, mean indices are constructed for stars in every field on each night and
compared to standard values.  With these transformation equations,
indices constructed from aperture photometry may be transformed to standard
values.  Because we determine the mean difference between the aperture
photometry and the profile-fit-based photometry for each index
for the program cluster stars on each photometric night, 
this calibration may be extended to the averaged profile-fit magnitudes and indices.

On two photometric nights of the January 2002 run, H$\beta$ indices were
obtained for stars in NGC 3680 as well as in the following open clusters with standardized
H$\beta$ photometry:  NGC 2287 \citep{GLH}, NGC 2516 \citep{Sno}; 
NGC 3766 \citep{Shob} and M 67 
\citep{NT87}. As the cluster standards cover a broad range
of H$\beta$ values, 2.57 to 2.93, they were used to establish the slope
of the calibration relation, leaving the determination of the zero point
to four field standards from the catalog of \citet{OLS83} observed
on one of the photmetric nights. The standard error of the mean 
zero-point for the calibration equation is 0.007.  NGC 3680
was observed with H$\beta$ filters on one of the photometric nights, 
with the mean difference between aperture- and profile-fit-based indices
determined with a standard error of the mean of 0.001 mag.

For the $hk$ indices, there are not yet very many open clusters with sufficient
standards to supplant the use of field star standards from the catalog
of \citet{TAT95}.  In all, 26 standards were observed on
three photometric nights for which $uvbyCa$ frames were obtained.  The
standards cover a range of over one magnitude in $hk$, so it was not difficult
to use these stars to establish both slope and intercept for the calibration
equation; a standard error of the mean of 0.005 mag is associated with the
zero-point of the calibration equation.  
NGC 3680 was observed on two of these photometric nights,
permitting determination of the mean difference between aperture and
profile-fit $hk$ indices with a precision of 0.004 mag (s.e.m.).

Open cluster stars and field stars observed on three photometric nights 
of the January 2002 run form the basis of the calibration to standard 
$uvby$ indices.  The open cluster standards
were drawn from stars observed by \citet{GLH} in NGC 2287, by
\citet{Shob} in NGC 3766, and by \citet{NT87} in M 67.
In addition, 23 field stars were observed over the three nights, several of
them giants with photometry from \citet{OLS93}.  Stars in common between
the nights among the field star standard sets and among the open 
clusters were used to map the three nights into a single homogeneous set.  
As NGC 3680 was also observed on two
of these photometric nights, it was straightforward to extend the calibrations
to the aperture photometry in NGC 3680, and further to the profile-fit
photometry.  

As before, the cluster standard sets were used to establish
the slopes of the relations between instrumental $V$, $(b-y)$, $m_1$ and
$c_1$, except that insufficient red giant stars were available in the
cluster standard sets to determine the distinct slopes and color terms for
the $m_1$ and $c_1$ relations; for these transformations, the cooler field
giants were exclusively used.  Finally, the 23 field stars were
entirely depended upon to set the zero-points of the calibration equations.
We estimate the standard errors of the mean zeropoints for the dwarf
calibration equations applied to NGC 3680 photometry
for $V$, $b-y$, $m_1$ and $c_1$ to be 0.006, 0.006,
0.009 and 0.010; comparable values for the cool giant calibration equations
are 0.010, 0.008, 0.010, and 0.017. The remaining contribution to the
zero-point uncertainty for the NGC 3680 photometry arises from the
determined mean difference between aperture-based indices and profile-fit-
based indices; comparisons from two nights on which NGC 3680 was observed
suggest that these corrections may be applied to the $V$, $b-y$, 
$m_1$ and $c_1$ indices with precisions of 0.004, 0.006, 0.008 and 0.008
mag respectively, based on the s.e.m. for the mean index differences.

Final photometric values are found in Table 2, where the primary identification and
coordinate description follow the WEBDA database conventions for this
cluster as of July 2003. Following each of the six photometric indices are the
standard errors of the mean for each index and a summary of the number of frames
in each of the seven bandpasses. Stars for which the $m_1$ and $c_1$ instrumental
indices were transformed using the giant calibration are indicated by brackets around
the indices. Stars with $V$ $\lesssim$ 11.8 or  $m_1$ $\gtrsim$ 0.25 were reduced as giants, as were
stars with $b-y$ $\gtrsim$ 0.46 and $m_1$ $\gtrsim$0.2.  These criteria were established
by examination of the known giants in the instrumental $m_1$, $c_1$ diagram.
Fig. 1 shows the standard errors of the mean for each index as a 
function of $V$.

\subsection{Comparison to Previous Photometry}
Though photometric indices on the $uvby$H$\beta$ system have been
published on a number of occasions for NGC 3680, the critical comparison
is with the data of \citet{NI88} since it has served as a source
of local standards for many of the surveys. Of the 33 stars with photoelectric
$uvby$ photometry, three (1,3,60) were either outside the limits of 
the survey or close enough to the edge to lack a sufficient number of 
frames in $b$ and/or $y$. Two of the stars (30,80) were excluded because
it has been demonstrated in past analyses that they are optical doubles
with anomalous indices as defined in \citet{NI88}. The residuals in the
comparisons between our data and that of \citet{NI88}, in the sense (ATT-NI),
are +0.014 $\pm$ 0.010, 0.007 $\pm$ 0.006, --0.019 $\pm$ 0.012, 
and --0.012 $\pm$ 0.019 for $V$, $b-y$, $m_1$, and $c_1$, respectively, from 
28 stars. The errors quoted are standard deviations for a single star. The 
residuals for H$\beta$, --0.008 $\pm$ 0.008, are based upon
only 8 stars. Generally, the offsets between the indices are modest, with
small scatter, particularly in $b-y$. The largest differential, however, is
that for $m_1$, with the indices of \citet{NI88} being 0.019 mag larger,
consistent with a larger implied metallicity; we will return to this point
in Sec. 4. The scatter in $c_1$ is somewhat
larger than expected, but is dominated by two of the fainter stars in the
sample near V $\sim$ 14.5. If these two stars are excluded, the 
scatter among the remaining 26 stars becomes 0.015 mag.

For $Caby$, the comparison between seven stars from Table 1 that overlap
with Table 2, in the sense (Table 2 - Table 1), produces residuals of
--0.004 $\pm$ 0.008, --0.001 $\pm$ 0.005, and +0.007 $\pm$ 0.006 for
$V$, $b-y$, and $hk$, respectively. The importance of this small
sample is that it is dominated by redder stars well beyond the color limits of
the \citet{NI88} survey and thus provides a check on the $V$ and $b-y$
data at the redder end of the scale, while testing the zero-point of the $hk$
indices. It is clear from the small offsets and the scatter in the residuals
that, particularly for the cooler stars, the CCD data are in very good
agreement with the standard system. If any correction were applied, the 
CCD $hk$ indices might be reduced by a few millimagnitudes. 

More extensive comparisons are available for $V$ and $b-y$ from three
CCD surveys. In the discussion that follows, we will make our comparisons
first using only stars brighter than $V$ = 15, then with all stars brighter than
$V$ = 17.0. The brighter stars are crucial because the majority
of the cluster members
lie above this cut and it is more effective to check for color terms among
the residuals if the photometry under discussion has small internal errors.

The survey by \citet{ATS89} (hereinafter referred to as ATS) 
was among the earliest attempts to
study clusters with intermediate-band photometry using CCD's and 
followed up on analyses of M67 \citep{AT87a}, NGC 6397 \citep{AT87b},
and IC 4651 \citep{ATT87}. Such programs were plagued by CCD's with
small fields of view, high readout noise, and low sensitivity in the blue and
ultraviolet, but still provided photometric indices of high internal precision.
Four fields were observed in NGC 3680 with only minimal overlap between
two of them. Though the $V$ and $b-y$ zero-points within the fields were
set using the data of \citet{NI88}, color terms for the entire $b-y$ range from
the turnoff to the giants were derived from external standards developed
within IC 4651, guaranteeing that the $b-y$ colors were on the standard
system, a point possibly missed by \citet{NO96} whose photometry suffers
from significant color terms.

Before the comparisons are made, corrections are in order for three stars.
The photometry listed for star 84 (ATS ID 2028) is actually that for 83 
(ATS ID 2027). The photometry published for 83 should undoubtedly be
that for 84, but the indices are clearly not plausible. It is likely that the
proximity of stars 83 and 84 is the source of the distortion. Likewise, the
indices for star 20 are incorrect and may be related to the fact that it is
the brightest star within the observed fields and the image suffered
from the effects of near saturation. Stars 20 and the correct 84 
will be excluded from the
comparison. As summarized in Table 3, for 43 stars with $V$ $\leq$ 15, the
mean differences in $b-y$ and $V$, in the sense (Table 2 - ATS), are
+0.007 $\pm$ 0.013 and +0.010 $\pm$ 0.013, in excellent agreement with
the comparison to \citet{NI88}, as expected.  If we expand the sample to all
stars $V \leq 17$, we need to exclude only star ATS 2054 
which exhibits large residuals in both $V$ and $b-y$. It should be 
emphasized that this star also
has exceptionally large error bars ($\sim$0.1 mag) for its apparent magnitude
in the original survey by \citet{ATS89}. For the remaining 79 stars, the
offsets in $b-y$ and $V$ are +0.004 $\pm$ 0.026 and +0.005 $\pm$0.023,
respectively. No color dependence is found among the residuals.

A more extensive survey with larger-format CCD's was carried out by
\citet{NO96}, who used only the photometry of \citet{NI88} to standardize
their data. As shown in Table 3, from 86 stars brighter than $V$ = 15,
the mean residuals in $b-y$ and $V$ are +0.017 $\pm$ 0.019 and +0.020
$\pm$ 0.022, respectively. However, closer examination reveals a significant
correlation of the residuals in $V$ and $b-y$ with $b-y$. If one corrects for 
this effect using the coefficients in Table 3, the dispersions in the residuals
reduce to $\pm$0.010 and $\pm$0.017 for $b-y$ and $V$, respectively.
For the sample with $V \leq 17$, two stars with excessive residuals,
511 and 527, are excluded. For the remaining 200 stars, the mean offsets
are +0.016 $\pm$0.023 and +0.029 $\pm$0.029 in $b-y$ and $V$, but
virtually identical color corrections emerge from the sample. With the color
terms included, the scatter reduces to $\pm$0.019 and $\pm$0.026.

The most recent attempt to use CCD intermediate-band photometry to
probe the nature of NGC 3680 is that of \citet{BR99}. From 61 stars brighter
than $V$ = 15, with no stars excluded, the offsets are +0.006 $\pm$0.013
and +0.014 $\pm$0.018 for $b-y$ and $V$, respectively. No color dependence
is found, as expected, since \citet{BR99} used standards over a wider
range of color to transform their indices. The color term in \citet{NO96} was
previously noted by \citet{BR99}. Expanding the analysis to all stars, with
no exclusions, the offsets become +0.005 $\pm$0.030 and +0.025 $\pm$0.032
for $b-y$ and $V$, respectively.

In summary, for the brighter stars, all three surveys exhibit impressively small
scatter in the residuals when compared with Table 2 for both $V$ and $b-y$.
After correcting for color terms, the $b-y$ data of \citet{NO96} are a tight match
to the current study, with \citet{ATS89} and \citet{BR99} only slightly worse.
The same trend generally applies for all stars with $V \leq 17$, though the
dispersions in the data of \citet{BR99} at the fainter end are noticeably 
worse than the other two studies.  

\section{The Color-Magnitude Diagram: Thinning the Herd}
Since one of the primary goals of this investigation is to test the 
analysis procedure for selecting probable single-star members
using a cluster with reliable membership information, we will treat 
the data for NGC 3680 as if the cluster were a
program object comparable to NGC 6253 (Paper III). This means that we will
defer making use of the considerable body of data bearing on the 
membership and binarity of stars in NGC 3680 until after selection on
photometric grounds has been attempted.

The CMD for all stars with at least 2 observations each in $b$ and $y$ 
is presented in Fig. 2. Open circles are the 182 stars with standard
errors in the mean for $b-y$ $\leq$ 0.010 mag. The morphology of the CMD is
identical to that found in the previous intermediate and broad-band CMD's:
the cluster is clearly of intermediate age with a well-defined red giant
clump and red hook at the main sequence turnoff. Below $V$ $\sim$ 15, the
identifiable main sequence becomes lost amid the increasing contamination
by field stars. 

To minimize potential contamination of the sample in the analysis of the
turnoff, our first cut will eliminate all stars with $V$ fainter than
15.0, $b-y$ greater than 0.40, and photometric uncertainty in 
$b-y$ greater than 0.011 mag. This reduces the sample to the 
45 stars plotted in Fig. 3. Fortunately, due to its galactic location and 
bright apparent turnoff, contamination of the restricted CMD by non-members 
should not be a significant problem for our analysis, so no attempt will be 
made to restrict the sample a second time by eliminating stars located 
well away from the cluster core, a procedure that proved invaluable 
in Paper III.

\subsection{Thinning the Herd: CMD Deviants}
Given the high precision of the $b-y$ indices and the expected dominance
of the cluster sample over the field stars, it is probable that the
majority of stars at the cluster turnoff are members, though not necessarily
single stars. Below the turnoff one can readily define the likely location
of the single-star main sequence, with two distinct groups of stars bracketing
the mean relation. The stars to the blue, noted as crosses in Fig. 3, are
tagged as probable field stars since no combination of single-star colors
in a composite system will move a star blueward of the main sequence CMD.
To the red, the band of stars above the main sequence (open triangles)
are tagged as probable binaries and/or nonmembers. Because of evolution off
the main sequence, the binary sequence at some point crosses and merges
with the vertical turnoff, making separation of the two impossible. The
bright limit on the identified binary sequence has been defined with 
this issue in mind.

Despite the high precision of the $b-y$ data, it is reasonable to question
some of the classifications of the stars that are located only 3 sigma off
the mean relation. Fortunately, the availability of multiple indices 
makes the selection process much less subjective. To demonstrate, 
we make use of the filter pair with the largest baseline, $u-y$. 
For this select sample of 45 stars, the standard error of the mean in $c_1$ 
is typically $\pm$0.007 mag and no star has an uncertainty larger than 
$\pm$0.011 mag. The comparable error in $u-y$ is less than this. 
The ($V$, $u-y$) CMD is shown in Fig. 4 and 
demonstrates the effectiveness of this index. The increased temperature
sensitivity of $u-y$ is obvious and allows easier identification of
stars that are too blue for their apparent magnitude (field stars) and
stars that are too red (binaries and/or field stars). The symbols have the
same meaning as in Fig. 3. With one exception, the stars that lie within the
binary sequence in the standard CMD have been identified as such from the
$u-y$ index. The one exception is easily explained and does not 
imply that the star is misidentified as a binary. The binaries that isolate themselves
significantly from the main sequence are composed of approximately similar
stars from the unevolved main sequence. The $u-y$ index for these stars
follows a simple relation between effective temperature and color. For
single stars in the vertical band at the turnoff, evolution off the main
sequence alters the energy distribution by decreasing the relative
contribution of the ultraviolet region blueward of the Balmer discontinuity.
Thus, $c_1$ increases as $M_V$ decreases at a given temperature. The
declining contribution of the $u$ filter leads to a redder $u-y$ index
due to surface gravity effects rather than temperature, making the
more evolved stars appear redder than less evolved stars at the same
temperature. The result is that the binary sequence will cross the
vertical turnoff at a redder and fainter location than in $b-y$. 
Note also that the $u-y$ data has identified one additional probable field
star that lies significantly redward of the main sequence in $u-y$ but only
slightly redward in $b-y$. This star is
likely to be a background giant or subgiant and will be excluded from the
analysis.

Elimination of the stars that scatter blueward in the $u-y$
figure would remove all the fainter scatter to the blue in the standard CMD.
Though it is always possible that elimination of these deviants 
from the analysis will sweep away a few true cluster members, 
the remaining sample will contain a high 
percentage of single stars whose structure and evolution are 
typical of the cluster as a whole. Finally, for consistency with past analyses, 
we remove two stars that
lie brighter and/or blueward of the main sequence turnoff. The positions
of these stars in the CMD would classify them as blue stragglers or
non-members. Since, by definition, the blue stragglers undergo
anomalous evolution, the inclusion of the parameters based upon
their indices is subject to question.

\section{Fundamental Properties: Reddening and Metallicity}
\subsection{Reddening}
From the sample of 35 stars selected as probable members 
from the filled circles in Figs. 3 and
4, we remove an additional star that has H$\beta$ indices based upon one
observation in one of the filters. For the remaining 34 stars, the standard
error of the mean in H$\beta$ is $\pm$0.004 mag with the largest
individual error at $\pm$0.012 mag. For $m_1$, the comparable statistics
are $\pm$0.006 mag for the typical error, with the largest individual 
value being $\pm$0.013 mag. 

As discussed in Paper I, derivation of the reddening from intermediate-band
photometry is a straightforward, iterative process given reliable estimates
of H$\beta$ for each star. The primary decision is the choice of the standard
relation for H$\beta$ versus $b-y$ and the adjustments required to correct
for metallicity and evolutionary state. The two most commonly used relations
are those of \citet{OLS88} and \citet{SN89}. As found in Paper I for
IC 4651, both
produce very similar but not identical results. We have processed the
indices for the 34 non-blue-straggler stars through both 
relations and find $E(b-y)$ = 0.040 $\pm$0.012 (s.d.) with \citet{OLS88} 
and $E(b-y)$ = 0.044 $\pm$0.011 (s.d.) with \citet{SN89}. As a compromise, 
we will take the weighted average of the two and use $E(b-y)$ = 0.042 
$\pm$0.002 (s.e.m.) or $E(B-V) = 0.058 \pm$0.003 (s.e.m.) in the 
analyses that follow.

\subsection{Metallicity from $m_1$} 

Given the reddening of $E(b-y)$ = 0.042, the derivation of [Fe/H] from
the $m_1$ index is as follows. The $m_1$ index for a 
star is compared to the standard relation at the same color and the
difference between them, adjusted for possible evolutionary effects, is a
measure of the relative metallicity. Though the comparison of $m_1$ is most
commonly done using $b-y$ as the reference color because it is simpler to
observe, the preferred reference index is H$\beta$ due to its insensitivity
to both reddening and metallicity. Changing the metallicity of a star will
shift its position in the $m_1$ - $(b-y)$ diagram diagonally, while moving
it solely in the vertical direction in $m_1$ - H$\beta$. Moreover, reddening
errors do not lead to correlated errors in both $m_1$ and H$\beta$. However,
as a check on the derived [Fe/H] based upon the H$\beta$-defined standard
relation, we will also derive the cluster abundance using $\delta$$m_1$ as
defined relative to a metallicity-adjusted $b-y$, which should produce
very similar results for stars near solar metallicity. This procedure is
identical to that applied to IC 4651 in Paper I.

The primary weakness of metallicity determination with intermediate-band filters
is the sensitivity of [Fe/H] to small changes in $m_1$; the typical slope of
the [Fe/H]/$\delta$$m_1$ relation is 11. Even with highly reliable photometry,
e.g., $m_1$ accurate to $\pm$0.015 for a faint star, the uncertainty in [Fe/H]
for an individual star is $\pm$0.17 dex from the scatter in $m_1$ alone.
When potential photometric scatter in H$\beta$ and $c_1$ are included,
errors at the level of $\pm$0.2 dex are common. As noted in 
previous papers in this series, the success of the technique 
depends upon both high internal accuracy 
and a large enough sample to bring the standard error of the
mean for a cluster down to statistically useful levels, i.e., below 0.10 dex.
Likewise, because of the size of the sample, we can also minimize the impact of 
individual points such as binaries and/or the
remaining non-members, though they will clearly add to the dispersion.

After correcting each star for the effect of $E(b-y)$ = 0.042, the
average $\delta$$m_1$ for 34 stars is +0.027 $\pm$0.002 (s.e.m.), which
translates into [Fe/H] = --0.175 $\pm$0.026 (s.e.m.) for our calibration.
For the modified $(b-y)$-based relation, the
average $\delta$$m_1$ is +0.026 $\pm$0.002 (s.e.m.), which 
translates into [Fe/H] = --0.168 $\pm$0.024.
The zero-point of the calibration has been set to match the
adopted value for the Hyades of +0.12. 

\subsection{Metallicity from $hk$}
We now turn to the alternative avenue for metallicity estimation,
the $hk$ index. The $hk$ index is based upon the addition 
of the $Ca$ filter to the traditional
Str\"{o}mgren filter set, where the $Ca$ filter is designed to measure the
bandpass which includes the H and K lines of Ca II. The design and development
of the $Caby$ system have been laid out in a series of papers discussing
the primary standards \citep{ATT91}, an extensive catalog of field star
observations \citep{TAT95}, and calibrations for both red giants \citep{ATT98}
and metal-deficient dwarfs \citep{ATT00}. Though the system was optimally
designed to work on metal-poor stars and most of its applications have
focused on these stars \citep{ATC95,BD96}, early indications that the system retained
its metallicity sensitivity for metal-rich F dwarfs have been confirmed
by observation of the Hyades and analysis of nearby field stars \citep{AT02}.

What makes the $hk$ index, defined as $(Ca-b)-(b-y)$, so useful even at the
metal-rich end of the scale is that it has half the sensitivity of $m_1$ to
reddening and approximately twice the sensitivity to metallicity changes,
i.e., $\delta$[Fe/H]/ $\delta$$hk$ = 5.6. The metallicity calibration for F stars
derived in \citet{AT02} used $\delta hk$ defined relative to $b-y$ as the
temperature index. To minimize the impact of reddening on metallicity, this
calibration was redone in Paper III
using H$\beta$ as the primary temperature index, leading to the relation

\medskip
\centerline{[Fe/H]$ = -3.51 \delta hk(\beta) + 0.12$}
\smallskip
\noindent
with a dispersion of only $\pm$0.09 dex about the mean relation. Though the
derived zero-point of the relation was found to be +0.07, it has been
adjusted to guarantee a Hyades value of +0.12, the same zero-point used for
the $m_1$ calibration. 

We have derived the cluster metallicity using the $hk$ indices for the same
34 stars analyzed above using both $hk$ relative to the $b-y$ relation and
$hk$ relative to H$\beta$. The results relative to $(b-y)$ and H$\beta$ are 
[Fe/H] = --0.137 $\pm$0.023 (s.e.m.) and --0.105 $\pm$0.016 (s.e.m.), respectively.

The unweighted average of the four determinations is [Fe/H] = --0.146 $\pm$0.032,
while inclusion of a weight based upon the inverse of the standard error of the
mean raises the average to [Fe/H] = --0.141. Given that the calibration curves for the $m_1$
and $hk$ metallicity relations and the photometric zero-points for the $m_1$ and $hk$
photometry were derived independently of each other, we find that the cluster
metallicity is reasonably well determined at [Fe/H] = --0.14 $\pm$0.03 (s.e.m.). 
The agreement between the abundances from $m_1$ and $hk$ is 
encouraging in that it is consistent with the revised zero-point for the $m_1$
photometry. An error of --0.02 mag in the mean $m_1$ values would require 
a correlated error of approximately --0.05 mag in the $hk$ zero-point to 
ensure that both systems produce similar estimates for [Fe/H]. It should
be noted also that in all four metallicity estimates, a significant 
portion of the dispersion in [Fe/H] arises from 3 of the fainter stars with 
systematically lower [Fe/H] than the cluster mean. Exclusion of these stars 
would increase [Fe/H] by only 0.03 dex, so they will be retained in the analysis.  

The impressive precision of the metallicity estimates as characterized 
by the standard errors of the mean is a product of the small to modest errors 
for individual stars coupled to a statistically significant sample of 
members and, thus, refers solely to the internal errors. 
On the external side, systematic errors may arise at a number of points along 
the way from processing of the frames to transformation of the instrumental 
indices to the standard system. Simple propagation of error analyses for 
the four techniques for defining [Fe/H] produce potential systematic 
uncertainties of $\pm 0.13$, 0.13, 0.10 and 0.06 dex from the 
$m_1(b-y)$, $m_1$(H$\beta$), $hk(b-y)$, and $hk$(H$\beta$) 
[Fe/H] determinations, 
respectively. Given the weighted average of the final values, 
the best estimate for the external error in the derived value of [Fe/H] is just under 0.10 dex.

\subsection{Comparison to Previous Determinations}
For the reddening, there are a significant number of determinations, though 
not all are independent. The first reddening estimate of
$E(B-V)$ = 0.04 by \citet{EG69} was based upon interpolation of the 
cluster distance relative to 3 early-type stars in the cluster field with 
$E(B-V)$ derived from $BV$ photoelectric photometry and spectral types. 
\citet{MC72} challenged the reliability of this approach and obtained $E(B-V)$ 
= 0.09 from DDO photometry of 8 red giant members of the cluster.  
\citet{JA79} revised the DDO estimate using an updated
photometric approach to obtain $E(B-V)$ = 0.06, but an additional
DDO survey by \citet{CL83}  found $E(B-V)$ = 0.10. The DDO
approach was revisited by \citet{TAT97} where the combined photometry
of \citet{MC72} and \citet{CL83} generated $E(B-V)$ = 0.095 with
the calibration of \citet{JA77}.  

A potential weakness of the early DDO results is that they required $B-V$
colors; regular use was made of the \citet{EG69} photoelectric data. 
Comparisons with more recent photolectric observations in the cluster 
\citep{AT91,KO97} indicate that the \citet{EG69} $B-V$ values for the 
giants may be 0.01 to 0.02 mag too red. Correction for this small shift 
in the relation used by \citet{MC72} would lower the DDO estimate by 
between 0.02 and 0.04 mag and the other estimates by slightly smaller 
values. It should be noted that the four stars classified as photometric 
non-members from the DDO data are now known to be non-members 
from both proper-motion and radial-velocity work, while all 8 stars included 
in the DDO analyses are members.

Among the giants, \citet{EG83} found $E(B-V)$ = 0.05 using a modified
Str\"{o}mgren system combined with $RI$ photometry, but later revised the
calibration to find $E(B-V)$ = 0.095 \citep{EG89}.
Though the issue of binarity in 3 of the 8 giants has been raised 
in the past, exclusion of these stars from the average has 
no significant statistical impact on the cluster mean.

For the turnoff stars, $uvby$H$\beta$ data imply 
$E(b-y)$ = 0.034 from 8 stars \citep{NI88}, 0.031 from 3 single-star 
members \citep{NO97}, and 0.048 from 11 single-star members \citep{BR99}.
The higher value of \citet{BR99} is surprising since their photometry 
is tied to the system of \citet{NI88} and they derive the reddening 
with the same relation. In all three cases, the derived [Fe/H] is 
approximately +0.1; lowering the metallicity by 0.2 dex should lead to 
a higher $E(b-y)$ by about 0.01 mag due to the expected bluer intrinsic 
colors. Such a correction would bring \citet{NI88} and \citet{NO97} into 
excellent agreement with the valued derived
in this investigation. Note that our slightly redder indices for both $(b-y)$ and
H$\beta$ relative to \citet{NI88} cancel out in the reddening determination.
 
Two more indirect estimates should be mentioned. \citet{KO97} optimized
a match between the cluster CMD and solar metallicity isochrones to
derive $E(B-V)$ = 0.075. The reddening maps of \citet{SC98} indicate
$E(B-V)$ = 0.094 in the direction of NGC 3680, an upper limit along this 
line of sight and surely higher than the cluster value. 
In summary, given the uncertainty inherent in all of these techniques, 
there appears to be no significant discrepancy with the value of
$E(B-V)$ = 0.06 derived from our $uvby$H$\beta$ photometry of 
the probable cluster members at the turnoff.

The metallicity estimates for the cluster have fallen into two
distinct camps. Prior $uvby$H$\beta$ photometric estimates at the turnoff have
been linked to the study by \citet{NI88} and have
generated numbers virtually identical with the original estimate of 
[Fe/H] = +0.09: +0.10 \citep{ATS89}, +0.11 \citep{NO97}, and +0.09
\citep{BR99}. In contrast, every recent analysis of the giants,
photometric and spectroscopic, has elicited values between --0.10
and --0.20. The homogenized DDO photometry of the member
giants has been discussed by \citet{TAT97} using the revised
calibration of \citet{TAT96} and gives [Fe/H] = --0.12; the result does 
not change if the binaries are excluded. The reddening adopted
in that discussion was $E(B-V)$ = 0.05. The moderate-dispersion
spectroscopic work of \citet{FJ93}, updated and expanded by 
\citet{FR02}, produces [Fe/H] = --0.19 for 6 member giants on
a scale that is typically 0.05 to 0.1 dex more metal-poor than the
revised DDO scale. The mean value remains unchanged if the
2 binary giants are excluded. The lower metallicity of the giants
has also been corroborated by a high-resolution spectroscopic
sample of 6 giants by \citet{PA01}, who find [Fe/H] = --0.17 on
a scale where the Hyades is +0.13; the mean remains unchanged
if the 2 binaries are excluded. Finally, Washington photometry of the 
giants by \citet{GE91} leads to
[Fe/H] = --0.14 on a system where the Hyades has [Fe/H] = +0.07.

In summary, there is now excellent agreement between the
spectroscopic and photometric metallicity
of the giants and that of the turnoff stars in NGC 3680 as determined
by both the $uvby$ and the $Caby$ systems. The source of the
long-standing discrepancy appears to be a modest zero-point offset
in the $m_1$ indices obtained by \citet{NI88}. We cannot overemphasize
the fact that this conclusion is not simply a choice of our zero-point over that
of  \citet{NI88}; it is always possible that our zero-point for $m_1$ is systematically
off by 0.01 mag. However,
the agreement between $m_1$ and the more sensitive $hk$ index, as well
as the extensive work on the giants, gives us confidence that the true [Fe/H]
of NGC 3680 is closer to --0.10 than +0.10.

\subsection{Comparison With Reality}
Before discussing the age and distance to the cluster from isochrone fits,
it is important to see if our approach to eliminating the deviants from the CMD
was successful or an exercise in bias. This can be accomplished because
the majority of the stars in our field have proper-motion 
\citep{KO95} and radial-velocity \citep{NO97}
data that allows us to determine if they are field stars and/or binaries.

At the faint end of the scale, three stars were removed from the CMD because
they were located blueward of the main sequence in both $b-y$ and $u-y$.
Stars 1037 and 2024 have only proper-motion data; both are 0\% probable
members. Star 2095 has both radial-velocity and proper-motion data; it is a 
non-member. On the red side of the main sequence, 5 stars were removed. 
Three (22,24,36) are members but spectroscopic binaries; two (16,3064) 
are non-members. Of the stars near the turnoff, excluded for being too blue 
and/or too bright,
both 57 and 48 are proper-motion and radial-velocity non-members.
Thus, every star identified as potentially being a field star or a binary does,
in fact, fall within one or both of these categories.

In contrast, of the 34 stars used in the derivation of the cluster parameters,
how many are probable non-members? Using the classification by \citet{NO97},
6 of the stars with proper motion and radial-velocity data are likely non-members:
29,31,82,1085,2093, and 3001. Of these, use of the $c_1$ index would have
quickly removed 29. It is located at the top of the turnoff, but has indices indicative
of a totally unevolved star. In fact, using the cluster reddening on this foreground
star places it below the main sequence in a $c_1$ - $(b-y)$ diagram. The other
five stars, however, have $c_1$ - $(b-y)$ indices that do not distinguish them
from the other cluster stars, single or binary. Surprisingly, the binary star members
follow a virtually identical distribution in the $\delta$$c_1$, $\delta$$M_V$
diagram as the single stars. Expectation would be that larger $c_1$ at a given
$(b-y)$ (or H$\beta$) would correlate with distance above the unevolved main
sequence for single stars; binaries would isolate themselves in such a figure
by having $c_1$ for an unevolved star but systematically larger $\delta$$M_V$, an
effect seen in analyses of NGC 752 \citep{TW83, DA94} and M67 \citep{NT87}
This does not appear to be the case in NGC 3680, despite the high internal
accuracy of the photometry.

It's worth noting that the fundamental cluster parameters of reddening
and metallicity derived with these six non-members included, are changed
only slightly if the non-members are explicitly excluded.  The 
derived reddening, for example, is only 0.001 mag larger with the six non-members excluded.
Derived estimates of [Fe/H] with the non-members excluded are lower by
$\sim 0.005$, except for the estimate based on $hk(b-y)$ which drops by 0.02.

We close this comparison by noting that of the six stars classed as non-members,
only two, 29 and 2093, are non-members in both radial velocity and proper motion.
Star 1085 is a radial-velocity member, but excluded on the basis of proper-motion.
The remaining three are 0 probability radial-velocity members, but moderate to
high probability proper motion members: 31 (69 \%), 82 (49 \%), and 3001 (70 \%).
Star 3001 is a virtual photometric twin for star 10, a similarity that extends to
the Li abundance. Star 31 is a photometric twin for 4002; both are classed as SB1
binaries. The Li for star 31 is also consistent with its location in the CMD. Though these
similarities may be pure statistical coincidence, it should be remembered that
NGC 3680 is a cluster in a state of ongoing dissolution. Based upon number counts
\citep{AT91}, proper motions \citep{KO95}, and radial velocities \citep{NO97}, it has
been known for some time that NGC 3680 suffers from a deficiency of lower main
sequence stars, i.e., those with $M_V$ greater than 5. Since there is every reason
to believe that the cluster formed with such stars, their absence is an indication that
dynamical evaporation has been a key process defining the current appearance of the
cluster. By extension, the process is no doubt continuing and, at any given point in
time, some stars that formed with the cluster should be in the transition state between
gravitationally bound and unbound. Thus, the similarities between stars that are
technically non-members and those that meet both velocity criteria may not be
purely coincidental. In short, inclusion of these three stars would have had no
significant impact on the determination of its parameters since they are virtual twins to
stars that are cluster members.

\section{Fundamental Parameters: Distance and Age}

\subsection{The Broad-Band CMD - Transforming the Data}
Given the reddening and metallicity, the traditional approach to deriving
the distance and age is to compare the cluster CMD to a 
set of theoretical isochrones for which the color
transformations and bolometric corrections between the theoretical 
and the observational plane 
reproduce the colors and absolute magnitudes of nearby stars with known
temperatures and abundances. From an age standpoint, the critical test
is whether or not a star of solar mass, composition, and age resembles
the sun. For open clusters of solar and subsolar abundance, the impact
of this issue on cluster ages and distances has been emphasized on a
variety of occasions \citep{TAT89,TAM,DA94,TAH,TAB}.

As we have noted in past analyses, isochrones 
are invariably created for broad-band systems and a check of the most
recent publications shows that the theoretical isochrones are available
on the $UBVRI$ system, among others, but not for $uvby$.
This problem has been solved in the past by making use of the fact
that $b-y$ is well correlated with $B-V$ at a given [Fe/H] with little
dependence on evolutionary state. In contrast with most clusters, however,
there is a wealth of $b-y$ CCD and photoelectric data for NGC 3680 but only
one precise CCD $BV$ survey. The early $BV$ work of \citet{EG69} has
been called into question when compared to more recent data \citep{ATS89,
NO96} and the photographic data of \citet{ATS89}, though consistent within a
zero-point shift for the stars $V \leq  15$, has larger photometric scatter
than one would like. Thus, the only reliable $BV$ study with high internal
precision is that of \citet{KO97}, which only includes independent
photometry for 143 probable members (data for 3 stars are taken from
other surveys.)

To maximize the internal accuracy, the five $by$ surveys to date, 
\citep{NI88, AT91, NO96, BR99} and this study, have been merged. The
photometry of Table 2 has been adopted as the standard system and
each sample, including the photoelectric $Caby$ data of Table 1, 
has been transformed to the standard system using the
offsets/linear relations calculated in Sec. 3.3. Stars noted as deviants in
each survey have been excluded unless otherwise noted below. All stars
with $V \leq 15$ have been included in the merger, though the data
from \citet{NI88, NO96} and the current investigation have been given twice
the weight of the remaining studies. For $V > 15$, only the
stars in \citet{NO96} and Table 2 have been merged.

To locate potential problems, all stars with standard deviations in $V$ and/or
$b-y$ greater
than 0.03 mag have been identified. For $V$, 14 stars meet the criteria, but
6 have $V$ $\geq$ 16.  The remaining 8 stars are 47, 51, 60, 84, 86, 2060,
4015, and 4104.  For $b-y$, only 9 stars meet the criteria, and 6 of these
are have $V$ $\geq$ 16. The remaining problems are 51, 60, and 2060.
With the exception of star 51, the errors in $V$, if real, would not seriously
impact the position of the star in the CMD while the errors in color would since
they translate into a $B-V$ shift of 0.07 to 0.10 mag. For the $V$ mag, we
can call upon three broad-band surveys to identify the probable source of
the problem: the CCD data of \citet{HAW, KO97}, and the photoelectric 
data of \citet{AT91}. For each of these studies, the mean offsets in $V$,
in the sense (Table 2 - ref), are found to be +0.033 $\pm$0.017 (s.d.) from
90 stars \citep{HAW}, +0.020 $\pm$0.012 from 82 stars \citep{KO97},
and --0.017 $\pm$0.019 from 19 stars \citep{AT91}. In each
comparison, stars that deviated from the mean residual by more than
0.05 mag were excluded from the mean. We note that past
comparisons to the photoelectric data of \citet{AT91} have often
exaggerated the uncertainty in this photometry by including the 
poorly calibrated CCD data with the brighter photoelectric data. It is
undoubtedly the case that the attempt to extend the photographic calibration
to fainter magnitudes using the CCD observations is a primary source of the
non-linear residuals seen in the fainter photographic photometry.

In the case of star 51, our CCD processing finds 2 stars at the expected position,
identified as 51A ($V$ = 15.16) and 51B ($V$ = 15.58) in Table 2. 
Most surveys find $V$ $\sim$ 14.65, though \citet{BR99} finds $V$ = 14.87. 
Because of the potential
composite nature of this star, it will be excluded from further discussion.
For 2060, Table 2 gives $(V,b-y)$ = (15.784,0.511) with no significant
variation in the indices. \citet{BR99} and \citet{NO96} find almost identical
values of (15.695,0.62) while \citet{HAW} find $V$ = 15.79. Since this star
is a non-member, it will be excluded from further discussion. For star 60 the
situation is more complicated since it is classified as a probable cluster
member. In $V$, there is no clear pattern of variation with 5 studies
providing an almost uniform distribution of values from 14.256 to 14.318;
we will adopt 14.29 as a compromise. The significant variation is in the
color. Using the $b-y,B-V$ transformations derived below, the $B-V$
index implied or directly observed for the star ranges from 0.536 \citep{BR99}
and 0.586 \citep{AT91} to 0.639 \citep{KO97} and 0.642 \citep{NI88}. As a
compromise, we will adopt $b-y$ = 0.40 and $B-V$ = 0.60 for the star,
noting that it may be a variable.

The next step in the procedure is the transformation from $b-y$ to $B-V$.
Using the $BV$ data of \citet{KO97} as the standard system, linear relations
were derived between $b-y$ and $B-V$, breaking the sample into blue and
red stars to account for the likelihood of a slope change between the two
color ranges.
For the red stars, the fit was also repeated using only stars brighter than
$V$ = 15.0, with no significant change found in the relation compared to the
entire sample.
The final adopted transformations are: 

\medskip
\centerline{$b-y \leq 0.32$}
\smallskip
\centerline{$B-V$ = 1.346($\pm$0.035)*$(b-y)$ + 0.025($\pm$0.011)  }
\medskip
\centerline{$(b-y) > 0.32$}
\smallskip
\centerline{$B-V$ = 1.777($\pm$0.014)*$(b-y)$ - 0.115($\pm$0.009)   }
\smallskip
\noindent
The standard deviations of the residuals about the mean relation 
for the regions are $\pm$0.008 for the 29 bluer stars and
$\pm$0.012 for the 46 redder stars $V \leq 15$.

To ensure optimal photometry on the standard $BV$ system for all the stars, 
particularly the giants, two additional comparisons were made. \citet{AT91}
published photoelectric data for 28 stars that were used in conjunction
with CCD frames to calibrate the photographic survey. Though the attempt
to extend the calibration to fainter stars proved flawed, the photoelectric data
were reasonably well tied to the standard $BV$ system. Comparing the
photoelectric data to the composite $by$ data transformed to $B-V$,
from 20 stars one finds the residuals in $V$ and $B-V$, in the sense (CCD-PE)
to be --0.017 $\pm$0.019 and 0.000 $\pm$0.025, respectively. If one
unusually deviant star, 4095, is excluded, the average residual in $B-V$ becomes
--0.004 $\pm$0.016. 

Photoelectric data on the Geneva system has also been published by
\citet{ME95} for 10 redder stars. Assuming the [B-V] listed for star 13 in 
their Table
1 is an error and should read 0.493, we have transformed the Geneva index
to $B-V$ using the relation provided by \citet{ME95}. The residuals in $V$
and $B-V$ from 9
stars relative to the transformed $by$ data, in the sense (CCD - ME), are
0.014 $\pm$0.014 and 0.005 $\pm$0.022, respectively. If the one star
exhibiting anomalous residuals, star 34, is excluded, the revised residuals
become 0.010 $\pm$0.005 and --0.001 $\pm$0.015, respectively.  

In addition to demonstrating that the $B-V$ data are on the standard system,
we can combine the results for star 53, the only giant not observed
in the $by$ sample, from the 3 surveys to minimize the impact of any
potential problem in the data of \citet{KO97}. Using the revised offsets, the
final $V$ and $B-V$ values for this star are found to be 10.879 $\pm$0.036
(s.d.) and 1.120 $\pm$0.012 (s.d.) from 3 measurements. The error in $V$ is
larger than one would like, but will have a negligible impact on the position of the
star in the CMD.

\subsection{The CMD Fit: NGC 752}
With the cluster $BV$ photometry available, the usual procedure would be
to compare the reddening-corrected CMD to a set of appropriate isochrones
and derive the age and distance. Instead, we will compare NGC 3680 to
the observational data for NGC 752. The rationale for this is straightforward.
With the revised metallicity for NGC 3680, the cluster is identical within the
uncertainties to NGC 752. The DDO data and moderate dispersion 
spectroscopy for NGC 752 combined give [Fe/H] = --0.09 from 16 giants 
\citep{TAT97}. The
turnoff photometry on the $uvby$ system, transformed to the system of
\citet{CB70}, gives [Fe/H] = --0.08 from 31 stars while the photometric
zero-point of \citet{TW83} gives --0.20 \citep{DA94}. The reddening 
estimate for NGC 752 from DDO and $uvby$ photometry
combined is $E(B-V)$ = 0.03 \citep{DA94}. By comparing NGC 3680 to a real
cluster, one avoids the problems inherent in choosing specific theoretical
isochrones tied to different color transformations and differences in the
adopted internal physics.

To superpose the clusters, we must apply differential shifts in color and 
apparent magnitude to account for the reddening differential and the difference
in the apparent modulus. For the color shift, we decrease the $B-V$ colors
of the stars in NGC 3680 by 0.03 mag. For the $\Delta$$V$, we use the
distribution of giants, in particular, the red giant clump defined by the
concentration of stars undergoing He-core burning prior to their second
ascent of the red giant branch. In NGC 752, the clump of stars is easily
identified at $V$ = 9.00 $\pm$0.05. For NGC 3680, the giant distribution
shows a strong concentration of stars between $V$ = 10.85 and 10.95; we
adopt 10.90 $\pm$0.05 as the clump location in NGC 3680, leading to
a differential of $\Delta$$V$ = --1.90. The resulting superposition of
the two CMD's is shown in Fig. 5. Note that we have included all stars
identified as members. Open
circles are stars in NGC 752, while asterisks are probable binaries. Filled
circles are stars in NGC 3680, with triangles representing the known binaries.
We also
emphasize that the majority of the photometry in NGC 752 is a composite
average of photoelectric data, though some of the fainter stars have solely
photographic estimates.

In Fig. 5, the agreement between the turnoff and unevolved main sequences
is excellent. The blue edges of the distribution for both clusters, populated
by single stars and binaries with mass ratios significantly less than 0.5,
superpose at a level such that a differential shift of more than 0.02 mag in 
$B-V$ in either direction would be readily detectable. The redder and fainter
turnoff for NGC 3680 demonstrates the greater age of the cluster. In 
contrast, the red giants, while matching
up well in $V$ by design, do not superpose at all in
$B-V$. The clump in NGC 3680 is almost 0.10 mag redder than in NGC 752.
It is impossible to simultaneously match the
luminosity and colors for the main sequences and the clumps of stars on the
giant branches of the two clusters. Since the primary difference between
the two clusters is the greater age of NGC 3680, one option may be that
the red giant clump undergoes a significant shift to the red between the
age of NGC 752 and NGC 3680. To investigate this option, we
turn to the theoretical isochrones.

\subsection{The CMD Fit: The Isochrones}
There is a variety of isochrone sets available for comparison with
broad-band photometry. For consistency with our previous discussion
of IC 4651 and because, when properly zeroed to the same color and
absolute magnitude scale, most sets produce similar results for
ages and distances, we will use the sets of \citet{GI02}
(hereinafter referred to as PAD). Moreover, based upon the many comparisons,
including our own \citep{AT91,DA94,ASH,TAH}, between open clusters 
and past and present generations of isochrones, we will only 
make use of isochrones that
include convective overshoot mixing. On a scale where solar metallicity is
Z = 0.019 and Y = 0.273, PAD isochrones were also obtained for (Y, Z)
 = (0.250, 0.008) and (0.300, 0.030) or [Fe/H] = --0.38 and +0.20,
respectively. Since isochrones with [Fe/H] = --0.10 to --0.15 are not
available, it was decided to define the cluster parameters assuming each
of the three abundances was appropriate and interpolate the results
for the derived [Fe/H].  As a first step, the solar metallicity isochrones
were checked to ensure that they were on our adopted color and 
absolute magnitude scale of
$M_V$ = 4.84 and $B-V$ = 0.65 for a solar mass star at 4.6 Gyrs. A
check of the solar metallicity isochrones of PAD shows that they are
too red by 0.032 mag in $B-V$ and too bright by 0.02 mag in $V$.
For consistency, these offsets have been applied to all the isochrones used,
though there is some evidence that a smaller correction may be more
appropriate for the more metal-poor isochrones \citep{TAB}. The
impact of such a differential correction on our conclusions is minor.

With $E(B-V)$ = 0.03 and 0.06 for NGC 752 and NGC 3680, respectively,
and NGC 3680 offset by --1.90 in $V$ relative to NGC 752, the 
cluster CMD's were compared to the isochrones
by deriving the apparent modulus for NGC 752 that guaranteed a match
to the main sequence between $V$ = 12.5 and 13.5. Fig. 6 shows 
an example of the fit for the solar metallicity isochrones with an apparent
modulus, (m-M), of 8.50 and ages of 1.26, 1.41, and 1.58 Gyrs.  For [Fe/H]
= +0.2, 0.0, and --0.38, the required apparent moduli are 8.80, 8.50, and
7.95, respectively. For NGC 752, the ages are 1.05, 1.30, and 2.15 Gyr while
for NGC 3680, the comparable numbers are 1.20, 1.55, and 2.55 Gyrs.
It should be emphasized that for the lowest [Fe/H], the agreement between
the observed and theoretical CMD's at the turnoff is less than ideal; we have
defined the ages based upon the color of the bluest point at the turnoff and
the location of the red hook relative to the isochrones. Rounding off to
the nearest 0.05, interpolation of the values for [Fe/H] between --0.10 and
--0.15 leads to an apparent modulus for NGC 752 of (m-M) = 8.30 $\pm$0.1 and,
for NGC 3680, (m-M) = 10.20.  For the ages, we have interpolated in log (age)
relative to [Fe/H]. For NGC 752 and NGC 3680, the
respective ages are 1.55 and 1.85 Gyr, with an approximate
uncertainty in both values near 0.1 Gyr.

For NGC 752, this age estimate is in excellent agreement with past values from
a variety of isochrone fits. \citet{DA94}, using the overshoot models of \citet{SCH}
found an age of 1.7 $\pm$0.1 Gyr. \citet{KO97}, using identical reddening and
distance modulus, found an age of 1.6 $\pm$0.2 Gyr using the Yale isochrones
\citep{GR87, CH92}. 
Given the differences in the models, the overall consistency is more than
satisfactory and implies that, when comparable cluster parameters are adopted,
the different sets of isochrones zeroed to the same scale 
should produce similar ages and moduli.
 
For NGC 3680, the comparisons are more discordant due to the frequent
adoption of
a higher metallicity for the cluster. \citet{KO97}, adopting $E(B-V)$ = 0.075 and
(m-M) = 10.45, find an optimum match for their overshoot parameter for an age
of 1.6 Gyr when compared to a solar composition isochrone; our match to the
solar composition models of PAD produces an apparent modulus of 10.40 and
an age of 1.55 Gyr. There is every reason to believe that had \citet{KO97} adopted
the correct metallicity and a slightly lower reddening, the Yale isochrones
would have produced an age and modulus comparable to 1.85 Gyr and 10.20.
Exactly the same argument can be made with the results of 
\citet{NO97} where, assuming $E(B-V)$ = 0.05 and Hyades metallicity, [Fe/H] =
+0.12, (m-M) = 10.65 and an age of 1.45 Gyr are derived using the isochrones of
\citet{SCH}. With a slightly higher reddening, our interpolated values for this
metallicity are (m-M) = 10.6 and 1.35 Gyr, effectively no difference.

We close this section by noting the results for another cluster, IC 4651, the
first cluster analyzed in Paper I. Unlike NGC 3680, there is little
disagreement regarding the metallicity of IC 4651 from either the main sequence
photometry \citep{AT00a} or the DDO photometry of the giants \citep{TAT97}: the
cluster is more metal-rich than the sun and essentially the same as the Hyades
within the uncertainties. From $uvby$H$\beta$ analysis, in Paper I we found
that $E(b-y)$ = 0.071 or 0.062, depending upon the choice of intrinsic
color relations. More important was the comparison between NGC 3680 and
IC 4651. On the assumption that the two clusters had the same [Fe/H], it
was found that the CMD's superposed exactly at the turnoff and on the
giant branch if $E(b-y)$ for IC 4651 was
larger than that of NGC 3680 by 0.04. Since we now know that NGC 3680 is
more metal-poor than IC 4651, the difference caused by reddening alone is
reduced to 0.03 mag in $E(b-y)$. With $E(b-y)$ = 0.042 for NGC 3680, this
implies $E(b-y)$ for IC 4651 of 0.07, thereby favoring the larger value but not
inconsistent with the smaller number, given the uncertainties.

Using the range in reddening values ($E(b-y)$ = 0.062 to 0.071), the 
age range and apparent modulus range derived in Paper I were 
1.7 to 1.6 $\pm$0.1 Gyr and 10.15 to 10.3 using the same source
of theoretical isochrones adopted above for NGC 3680. The isochrones 
were transformed directly from the theoretical plane to $b-y$ for a 
Hyades composition and a direct fit made to the Hyades 
$V$ - $(b-y)$ main sequence. The age and distance were in
excellent agreement with comparable prior
analyses based upon the same metallicity and reddening when use was
made of isochrones with convective overshoot. Since then, \citet{ME00}
adopted $E(b-y)$ = 0.076 and fit the $V$ - $(v-y)$ relation directly 
to the Hyades to get (m-M) = 10.36, identical
with the values of Paper I if the same reddenings are selected. 

The age values have been corroborated with the same reddening but independent 
$uvby$ photometry by \citet{ME02} and the isochrones of \citet{YY} and \citet{SCH}.
 For $E(b-y)$ = 0.062 to 0.071, the isochrones of \citet{SCH} give 1.7 to 1.6 Gyr,
while the \citet{YY} isochrones generate ages systematically larger by 0.2 Gyr.
It should be noted that the age estimates in \citet{ME02} are heavily weighted by
the few points identified as single stars from radial-velocity work and that the
turnoff region in $b-y$ shows somewhat larger scatter than found in Paper I.
The flexible and somewhat subjective nature of the fitting procedure, compounded
with the differences in the models, may account
for the slightly higher age of the \citet{YY} fit and the systematically lower moduli
between (m-M) = 10.00 and 10.10.

To summarize, when analyzed in an internally consistent manner, the clusters 
NGC 752, IC 4651, and NGC 3680 have $E(B-V)$ = 0.03, 0.10, and 0.06,
respectively. The metallicities of NGC 752 and NGC 3680 are effectively
the same and lie between [Fe/H] = --0.10 and --0.15, on a scale where the
Hyades and IC 4651 have [Fe/H] = +0.12. The respective apparent moduli
for NGC 752, IC 4651, and NGC 3680 are 8.3, 10.2, and 10.2. Finally, the
correct age sequence for the clusters, with the revised [Fe/H] for NGC 3680,
is 1.55, 1.7, and 1.85 Gyr, respectively, with a typical uncertainty of $\pm$0.1
Gyr. It should be noted that while one may quarrel with the absolute scale
because of the choice of a particular set of isochrones, the relative rankings
should be immune to changes other than in the reddening and/or the 
metallicity. Given these parameters, we now turn our attention to the giant
branch.

\subsection{The CMD: The Giant Branch and Li}  
While the turnoff regions and the luminosity of the clumps for NGC 752 and 
NGC 3680 are quite compatible, one is faced with the color discrepancy in
the red giant clump: the clump for NGC 3680 is almost 0.10 mag redder
than in NGC 752. While theory predicts that the clump should generally shift
toward the red with age for intermediate age clusters, the theoretical isochrones
do not predict a change as large as that evidenced by NGC 752 and NGC 3680
for an age change of 0.3 Gyr \citep{GI00}. 
However, the color difference is consistent with
the shift between stars in the clump and the first-ascent red giant branch. In fact,
the reddest star at clump level in NGC 752, usually assumed to be a first-ascent
giant, is superposed within the clump of NGC 3680, while the bluer star in the
clump of NGC 3680 is surrounded by the clump of NGC 752. 

One interpretation is that the red giant region of NGC 752 
is dominated by He-core-burning stars
while that of NGC 3680 is dominated by first-ascent red giants. This isn't the
first time this suggestion has been made. \citet{PA01} have measured Li
abundances for an array of stars in NGC 3680 ranging from the tip of the giant
branch to the fainter regions of the turnoff. The surprising result was that the majority
of the stars in the clump (three out of four) have measurable Li  with log(Li) + 12
near 1.0. In contrast, the clump stars in NGC 752 have only upper limits at
a factor
of 5 or more below those in NGC 3680 and the stars classed as first-ascent
giants in NGC 752. The effect is illustrated in Fig. 7 where we have plotted
the Li abundance as a function of $V$ for NGC 752 and NGC 3680 with the
$V$ magnitudes for NGC 3680 reduced by 1.90 to place them on the same scale.
Upper limits are plotted as triangles, filled symbols refer to NGC 3680, open symbols
to NGC 752. All stars, binaries or not, have been included in the figure. The
bifurcation among the stars in the giant branch clump is relatively apparent. Note also
the respectable match in $V$ for the Li-dips on the main sequence. What is
new in the analysis is the fact that the Li-bifurcation superposes on the color
distribution of the giants, a fact previously lost due to the uncertainty in the
metallicity of NGC 3680 relative to NGC 752. A significantly more metal-rich
NGC 3680 would be expected to have a redder clump than NGC 752. With
similar metallicity, the luminosity and color distribution should be almost
identical.

In Fig. 8, we show the composite giant branch for NGC 752
(circles), IC 4651 (triangles), and NGC
3680 (squares). The photometry has been corrected for reddening 
and distance and the
points for IC 4651 have been reduced in color by an additional 0.01 mag  in
$b-y$ to
partially correct for metallicity effects. The color index chosen for the temperature
scale is $(b-y)$ due to its lower sensitivity to metallicity changes. 
As the clusters age, the color distribution of the stars near the clump shifts from
primarily blue, mixed range of blue to red, to mostly red in NGC 3680.
Though the majority of these stars have been analyzed for Li, only the filled
symbols identify stars with measured [log(Li) + 12] of +1.0 or higher. With the
exception of the fourth clump star in NGC 3680 with a Li abundance of +0.3, all
other determinations are upper bounds. Of the 6 stars with definitive high values, three
fall at the clump level in NGC 3680; the remaining three are spread out along the
apparent path of the first-ascent giant branch.

If the giants in NGC 3680 are predominantly first-ascent giants, why do they
simulate a clump? The distribution of first-ascent giants is partially controlled by the
bump that results when the H-burning shell crosses the discontinuity in 
composition created by the deep limit of the convective atmosphere as the
star ascends the giant branch. This discontinuity causes the star to retreat back
down the giant branch and then reascend, creating a bump in the 
distribution due to the increased length of time spent in this luminosity range.
Examples of this effect are seen in the giant tracks of Fig. 6. What is
important to note is that the low-luminosity limit of the bump grows fainter
as the mass of the star ascending the giant branch declines. For clusters
younger than NGC 752, the bump stars are expected to lie above the
level of the red giant clump; for clumps the age of NGC 3680 or higher, the
bump approaches the same level as the clump created by the He-core
burning stars. Thus, the possibility exists that the clump in NGC 3680
represents a totally different phase of evolution compared with the stars
in NGC 752.

Why should the ratio of first-ascent to second-ascent giants change so
dramatically? The only option we can offer comes from the one key
parameter that changes as the clusters age from NGC 752 to NGC 3680, the drop
in stellar mass. The stars that populate the giant branch are at the critical mass
range where He-ignition changes from a quiescent phase to degenerate-core
flash \citep{GI00}. The implication is that the development of 
the degenerate core and the
evolution after ignition alter the luminosity function in a way that minimizes the
He-core burning clump while coincidentally enhancing the bump at a luminosity
comparable to the clump. The only other alternative that presents itself is that
these relatively Li-rich stars are second-ascent giants, but the changeover in the
internal structure severely reduces the rate of Li 
destruction in the post-main-sequence
phase. The evidence available to date is inadequate to decide if either of these
two options is correct. At minimum, it appears certain that standard stellar
evolution does not predict this behavior among the giants in this age range.
Because of the small number statistics, comprehensive observations of
potential indicators of evolutionary states among giants, e.g., Li and C isotope
ratios, for every cluster giant member of a number of clusters of very
comparable age may be required ultimately to answer this
new riddle.
 
\section{Summary and Conclusions}                                     
Our understanding of NGC 3680 in the context of constraining stellar and galactic
evolution has followed an uneven path, with grudging progress as successive
studies have revised, clarified and added to previous work while leaving 
key questions unresolved. NGC 3680 ranked with NGC 752 
\citep{TW83} as one of the first clusters classified as having what was 
then described as a bimodal main sequence \citep{NI88}. The work of
\citet{NI88} and \citet{ATS89} refined the structure of the turnoff and
collectively demonstrated via $uvby$ photometry that the stars redward
of the vertical turnoff in the CMD were probable members and 
unlikely binaries, identical to
the pattern found in NGC 752. Recent work, including \citet{NO96,BR99} and
this study, has confirmed that this photometric spread in $b-y$ is real. 
Since standard stellar evolution models were incapable of producing single
or binary stars with the appropriate redder colors at the turnoff, an alternative
mechanism was necessary. 

The suggestion that core convective overshoot
could explain the peculiar CMD structure in NGC 3680 was first made by
\citet{MP88} and explicitly tested by \citet{ATS89} using the preliminary
overshoot models of  \citet{BE}. It was concluded that the red hook was 
naturally produced by single stars undergoing convective overshoot 
convolved with the normal
binary sequence approximately 0.7 mag above the main sequence.
Though the early attempts at creating isochrones with convective overshoot
for stars in the intermediate-mass ranges were flawed,
confirmation of this solution came through comprehensive identification of
single-star members and binaries in an open cluster with the
study of NGC 752 by \citet{DA94} and the revised isochrones of
\citet{SCH}. The same issue for NGC 3680 has 
been investigated periodically  \citep{AN90, CA93, NO97, KO97} as 
stellar models and the observational
data for the cluster have improved. Despite the sparse population of the
cluster and the rich population of binaries, the predominant debate that
remains is not the existence of convective overshoot, but the required size
of the phenomenon and its dependence on other factors such as
metallicity.  The necessity for the phenomenon again
lies with the stars that define the red hook at the turnoff; 
proper motions and radial velocities
have confirmed their original photometric classification as single-star members.
The small population of these stars, however, limits the definition of the
amount of overshoot unless the entire CMD is used, including the red giants,
and the sample is enhanced by merger with clusters similar in age and
metallicity, as in NGC 752 and IC 4651.

Unfortunately, for reasons discussed earlier, definitive values for the age were
unattainable due to the controversy over the true metallicity. Estimates of the
effect on the age of changing [Fe/H] between the two camps were typically
25 \% \citep{PA01}, though \citet{KO97} split the difference by adopting a
solar metallicity. With the metallicity issue resolved and the agreement between
the giant branch and the turnoff confirmed, NGC 3680 has been placed 
on an internally consistent reddening, metallicity, age, and distance scale with
NGC 752 and IC 4651. The combination of membership information with
high quality photometry demonstrates that despite a modest difference in age
and turnoff mass, the stars that define the clump in NGC 3680 are systematically
redder than those in NGC 752, while the giants in IC 4651 appear to straddle
both camps. The limited data from Li are consistent with the claim that the
so-called clump stars in NGC 3680 are, in fact, likely to be first-ascent red
giants, as originally suggested by \citet{PA01}. Whether this suggestion holds
up under greater scrutiny, only time and more definitive observational data will
tell. However, as with the early discrepancies between observation and
theory noted prior to the inclusion of convective overshoot, the cluster data are
indicative of a real effect not predicted by the current standard models for
post-main-sequence evolution.     

\acknowledgements
We should like to acknowledge the helpful and constructive comments made
by the referee. 
The progress in this project would not have been possible without the time
made available by the TAC and the invariably excellent support
provided by the staff at CTIO. The majority of the paper was completed
during an extended visit to CTIO and the authors gratefully acknowledge
the use of the facilities and the support received from the staff during
this period. Suchitra Balachandran kindly communicated her results on Li 
in IC 4651 in advance of publication. Extensive use was made of the SIMBAD
database, operating at CDS, Strasbourg, France  and the WEBDA database
maintained at the University of Geneva, Switzerland.
The cluster project has been helped by support supplied through 
the General Research Fund of
the University of Kansas and from the Department of Physics and
Astronomy.

\clearpage
\figcaption[fig1.eps]{Standard errors of the mean for $V$, and all
color indices as a function of $V$. Major tickmarks for the y axis are
0.05 apart.  The vertical scale for each successive index has been offset
by 0.05 mag for visual clarity.\label{fig1}} 

\figcaption[fig2.eps]{Color-magnitude diagram for stars with at least
2 observations each in $b$ and $y$. Crosses are stars with
internal errors in $b-y$ greater than 0.010 mag. \label{fig2}}

\figcaption[fig3.eps]{CMD for stars at the turnoff with errors in $b-y$
below 0.011 mag. Crosses are stars selected as field stars blueward
of the main sequence while triangles are probable field stars and/or
binaries above the main sequence. \label{fig3}}

\figcaption[fig4.eps]{CMD for the same stars as Fig. 3, using $u-y$
as the temperature index. \label{fig4}}  

\figcaption[fig5.eps]{CMD of all potential members of NGC 3680 superposed
on the CMD for NGC 752. Filled triangles are probable binaries in NGC 3680
while filled circles are all other stars. Asterisks are probable binaries in NGC 752
while open circles are all other stars. The data in NGC 3680 is shifted by -0.03
in $B-V$ and -1.90 in V. \label{fig5}}

\figcaption[fig6.eps]{Same data as Fig. 5 with binaries in NGC 752 removed and
all stars in NGC 3680 drawn as filled circles. Isochrones have solar abundance
and have been corrected for reddening and distance by $E(B-V)$ = 0.03 and
$(m-M)$ = 8.50. \label{fig6}}

\figcaption[fig7.eps]{Li abundance measures in NGC 752 (open symbols) and
NGC 3680 (filled symbols) as a function of $V$ with NGC 3680 at the same
distance and reddening as NGC 752. Triangles are upper limits. \label{fig7}}

\figcaption[fig8.eps]{Composite absolute CMD for all the giants in NGC 752 (circles),
IC 4651 (triangles), and NGC 3680 (squares). Filled symbols are all stars
with measured log(Li) + 12 above +1.0. \label{fig8}} 


\begin{thebibliography}{}
\bibitem[Andersen et al. (1990)]{AN90} Andersen, J., Nordstr\"{o}m, B., \&
Clausen, J. 1990, \apj, 363, L33
\bibitem[Andrievsky et al. (2002a)]{AN12} Andrievsky, S. M., Bersier, D.,
Kovtyukh, V. V., Luck, R. E., Maciel, W. J., L\'{e}pine, J. R. D., \&
Beletsky, Yu. V. 2002a, \aap, 384, 140
\bibitem[Andrievsky et al. (2002b)]{AN02} Andrievsky, S. M., Kovtyukh, V. V.,
Luck, R. E., L\'{e}pine, J. R. D., Maciel, W. J., \& Beletsky, Yu. V. 2002b,
\aap, 392, 491
\bibitem[Anthony-Twarog (1987a)]{AT87a} Anthony-Twarog, B. J. 1987a, \aj, 93,
647 
\bibitem[Anthony-Twarog (1987b)]{AT87b} Anthony-Twarog, B. J. 1987b, \aj, 93,
1454
\bibitem[Anthony-Twarog et al. (1991a)]{AT91} Anthony-Twarog, B. J., Heim, E. A.,
Twarog, B. A., \& Caldwell, N. 1991a, \aj, 102, 1056
\bibitem[Anthony-Twarog et al. (1991b)]{ATT91} Anthony-Twarog, B. J., Laird, J. B., Payne, D., 
\& Twarog, B. A. 1991b, \aj, 101, 1922
\bibitem[Anthony-Twarog et al. (2000)]{ATT00} Anthony-Twarog, B. J., Sarajedini, A., Twarog, 
B. A., \& Beers, T. C. 2000, \aj, 119, 2882
\bibitem[Anthony-Twarog \& Twarog (1987)]{ATT87} Anthony-Twarog, B. J., \& Twarog,
B. A. 1987, \aj, 94, 1222
\bibitem[Anthony-Twarog \& Twarog (1998)]{ATT98} Anthony-Twarog, B. J., \& Twarog, B. A., 
1998, \aj, 116, 1902
\bibitem[Anthony-Twarog \& Twarog (2000a)]{AT00a} Anthony-Twarog, B. J., \& Twarog,
B. A. 2000a, \aj, 119, 2282 (Paper I)
\bibitem[Anthony-Twarog \& Twarog (2000b)]{AT00b} Anthony-Twarog, B. J., \& Twarog,
B. A. 2000b, \aj, 120, 3111 (Paper II)
\bibitem[Anthony-Twarog et al. (1995)]{ATC95} Anthony-Twarog, B. J., 
Twarog, B. A., \& Craig, J. 1995, \pasp, 107, 32
\bibitem[Anthony-Twarog et al. (1994)]{ASH} Anthony-Twarog, B. J., Twarog, B. A., \&
Sheeran, M. 1994, \pasp, 106, 486
\bibitem[Anthony-Twarog et al. (1989)]{ATS89} Anthony-Twarog, B. J., Twarog, B. A.,
\& Shodhan, S. 1989, \aj, 98, 1634
\bibitem[Anthony-Twarog et al. (2002)]{AT02} Anthony-Twarog, B. J., Twarog, B. A.,
\& Yu, J. 2002, \aj, 124, 389
\bibitem[Baird (1996)]{BD96} Baird, S. R. 1996, \aj, 112, 2132
\bibitem[Bertelli et al. (1988)]{BE} Bertelli, G., Bressan, A. G., Chiosi, C., Nasi, E.,
\& Pigatto, L. 1988, preprint 
\bibitem[Bruntt et al. (1999)]{BR99} Bruntt, H., Frandsen, S., Kjeldsen, H.,
\& Andersen, M. I. 1999, \aaps, 140, 135
\bibitem[Carraro et al. (1993)]{CA93} Carraro, G., Bertelli, G., Bressan, A., \&
Chiosi, C. 1993, \aaps, 101, 381
\bibitem[Chaboyer et al. (1992)]{CH92} Chaboyer, B., Deliyannis, C. P., Demarque,
P., Pinsonneault, M. H., \& Sarajedini, A. 1992, \apj, 388, 372
\bibitem[Clari\'{a} \& Lapasset (1983)]{CL83} Clari\'{a}, J. J., \& Lapasset, E. 
1983, JApA, 4, 117
\bibitem[Corder \& Twarog (2001)]{SC01} Corder, S., \& Twarog, B. A. 2001, 
\aj, 122, 895
\bibitem[Crawford (1975)]{CR75} Crawford, D. L. 1975, \aj, 80, 955
\bibitem[Crawford \& Barnes (1970)]{CB70} Crawford, D. L., \& Barnes, J. V. 1970,
\aj, 75, 946
\bibitem[Daniel et al. (1994)]{DA94} Daniel, S. A., Latham, D. W., 
Mathieu, R. D., \& Twarog, B. A. 1994, \pasp, 106, 281 
\bibitem[Edvardsson et al. (1993)]{ED93} Edvardsson, B., Andersen, J., 
Gustaffson, B., Lambert, D. L., Nissen, P. E., \& Tomkin, J. 1993, \aap,
275, 101
\bibitem[Eggen (1969)]{EG69} Eggen, O. J. 1969, \apj, 155, 439
\bibitem[Eggen (1983)]{EG83} Eggen, O. J. 1983, \aj, 88, 813
\bibitem[Eggen (1989)]{EG89} Eggen, O. J. 1989, \pasp, 101, 366
\bibitem[Friel \& Janes (1993)]{FJ93} Friel, E. D., \& Janes, K. A. 1993,
\aap, 267, 75
\bibitem[Friel et al. (2002)]{FR02} Friel, E. D., Janes, K. A., Tavarez, M., Scott,
J., Hong, L., \& Miller, N. 2002, \aj, 124, 2693
\bibitem[Geisler et al. (1991)]{GE91} Geisler, D., Claria, J. J., \& Minniti, D.
1991, \aj, 102, 1836
\bibitem[Girardi et al. (2000)]{GI00} Girardi, L., Mermilliod, J. -C., \& Carraro, G.
2000, \aap, 354, 892
\bibitem[Girardi et al. (2002)]{GI02} Girardi, L. et al. 2002, \aap, 391, 195
\bibitem[Green et al. (1987)]{GR87} Green, E. M., Demarque, P., \& King, C. R.
1987, Revised Yale Isochrones and Luminosity Functions (Yale University
Observatory, New Haven)
\bibitem[Harris et al. (1993)]{GLH} Harris, G. L. H., FitzGerald, M. P. V.,
Mehta, S. \& Reed, B. C. 1993, \aj, 106, 1533
\bibitem[Hawley et al. (1999)]{HAW} Hawley, S. L., Tourtellot, J. G., \& Reid, I. N. 
1999, \aj, 117, 1341
\bibitem[Janes (1977)]{JA77} Janes, K. A. 1977, \aj, 82, 35
\bibitem[Janes (1979)]{JA79} Janes, K. A. 1979, \apjs, 39, 135 
\bibitem[Kozhurina-Platais et al. (1997)]{KO97} Kozhurina-Platais, V., 
Demarque, P., Platais, I., Orosz, J. A., \& Barnes, S. 1997, \aj, 113, 1045
\bibitem[Kozhurina-Platais et al. (1995)]{KO95} Kozhurina-Platais, V.,
Girard, M., Platais, I., \& Van Altena, W. F. 1995, \aj, 109, 672
\bibitem[L\'{e}pine et al. (2003)]{LP03} L\'{e}pine, J. R. D., Acharova, I. A., \& Mishurov,
Yu. N. 2003, \apj, 589, 210
\bibitem[Luck et al. (2003)]{LU03} Luck, R. E., Gieren, W. P., Andrievsky, S. M.,
Kovtyukh, V. V., Fouqu\'{e}, P., Pont, F., \& Kienzle, F. 2003, \aap, 401, 939 
\bibitem[Mazzei \& Pigatto (1988)]{MP88} Mazzei, P., \& Pigatto, L. 1988, \aap,
193, 148 
\bibitem[McClure (1972)]{MC72} McClure, R. D. 1972, \apj, 172, 615
\bibitem[Meibom (2000)]{ME00} Meibom, S. 2000, \aap, 361,929
\bibitem[Meibom et al. (2002)]{ME02} Meibom, S., Andersen, J., \& Nordstr\"{o}m
2002, \aap, 386, 187
\bibitem[Mermilliod et al. (1995)]{ME95} Mermilliod, J. -C., Andersen, J.,
Nordstr\"{o}m, B., \& Mayor, M. 1995, \aap, 299, 53
\bibitem[Mishurov et al. (2002)]{MI02} Mishurov, Yu. N., L\'{e}pine, J. R. D., \& Acharova,
I. A. 2002, \apj, 571, L113
\bibitem[Nissen (1988)]{NI88} Nissen, P. E. 1988, \aap, 199, 146
\bibitem[Nissen et al. (1987)]{NT87} Nissen, P. E., Twarog, B. A., \& Crawford,
D. L. 1987, \aj, 93, 634
\bibitem[Nordstr\"{o}m et al. (1996)]{NO96} Nordstr\"{o}m, B., Andersen, J., 
\& Andersen, M. I. 1996, \aaps, 118, 407
\bibitem[Nordstr\"{o}m et al. (1997)]{NO97} Nordstr\"{o}m, B., Andersen, J., \& Andersen,
M. I. 1997, \aap, 322, 460
\bibitem[Olsen (1983)]{OLS83} Olsen, E. H. 1983, \aaps, 54, 55
\bibitem[Olsen (1988)]{OLS88} Olsen, E. H. 1988, \aap, 189, 173
\bibitem[Olsen (1993)]{OLS93} Olsen, E. H. 1993, \aaps, 102, 89
\bibitem[Olsen (1994)]{OLS94} Olsen, E. H. 1994, \aaps, 106, 257
\bibitem[Pasquini et al. (2001)]{PA01} Pasquini, L., Randich, S., \& Pallavicini, R.
2001, \aap, 375, 851
\bibitem[Schaller et al. (1992)]{SCH} Schaller, G., Schaerer, D., Meynet, G., \&
Maeder, G. 1992, \aaps, 96, 269
\bibitem[Schlegel et al. (1998)]{SC98} Schlegel, D. J., Finkbeiner, D. P.,
\& Davis, M. 1998, \apj, 500, 525
\bibitem[Schuster \& Nissen (1989)]{SN89} Schuster, W. J., \& Nissen, P. E. 1989, \aap, 221, 
65
\bibitem[Shobbrook (1985)]{Shob} Shobbrook, R. R. 1985, \mnras, 212, 591
\bibitem[Snowden (1975)]{Sno} Snowden, M. S. 1975, \pasp, 87, 721
\bibitem[Twarog (1983)]{TW83} Twarog, B. A. 1983, \apj, 267, 207
\bibitem[Twarog \& Anthony-Twarog (1989)]{TAT89} Twarog, B. A., \& Anthony-Twarog 1989,
\aj, 97, 759
\bibitem[Twarog \& Anthony-Twarog (1995)]{TAT95} Twarog, B. A., \& Anthony-Twarog, B. J. 1995, 
\aj, 109, 2828
\bibitem[Twarog \& Anthony-Twarog (1996)]{TAT96} Twarog, B. A., \& Anthony-Twarog, B. J. 1996, 
\aj, 112, 1500
\bibitem[Twarog et al. (1999)]{TAB} Twarog, B. A., Anthony-Twarog, B. J., \&
Bricker, A. R. 1999, \aj, 117, 1816
\bibitem[Twarog et al. (2003)]{TW03} Twarog, B. A., Anthony-Twarog, B. J., \&
De Lee, N. M. 2003, \aj, 125, 1383 (Paper III)
\bibitem[Twarog et al. (1995)]{TAH} Twarog, B. A., Anthony-Twarog, B. J., \&
Hawarden, T. G. 1995, \pasp, 107, 1215
\bibitem[Twarog et al. (1993)]{TAM} Twarog, B. A., Anthony-Twarog, B. J., \&
McClure, R. D. 1993, \pasp, 105, 78 
\bibitem[Twarog et al. (1997)]{TAT97} Twarog, B. A., Ashman, K. M., \& Anthony-Twarog, B. A. 
1997, \aj, 114, 2556
\bibitem[Yi et al. (2002)]{YY} Yi, S. K., Kim, Y. -C, \& Demarque, P. 2002, \apjs, 144, 259

\end{thebibliography}
\enddocument
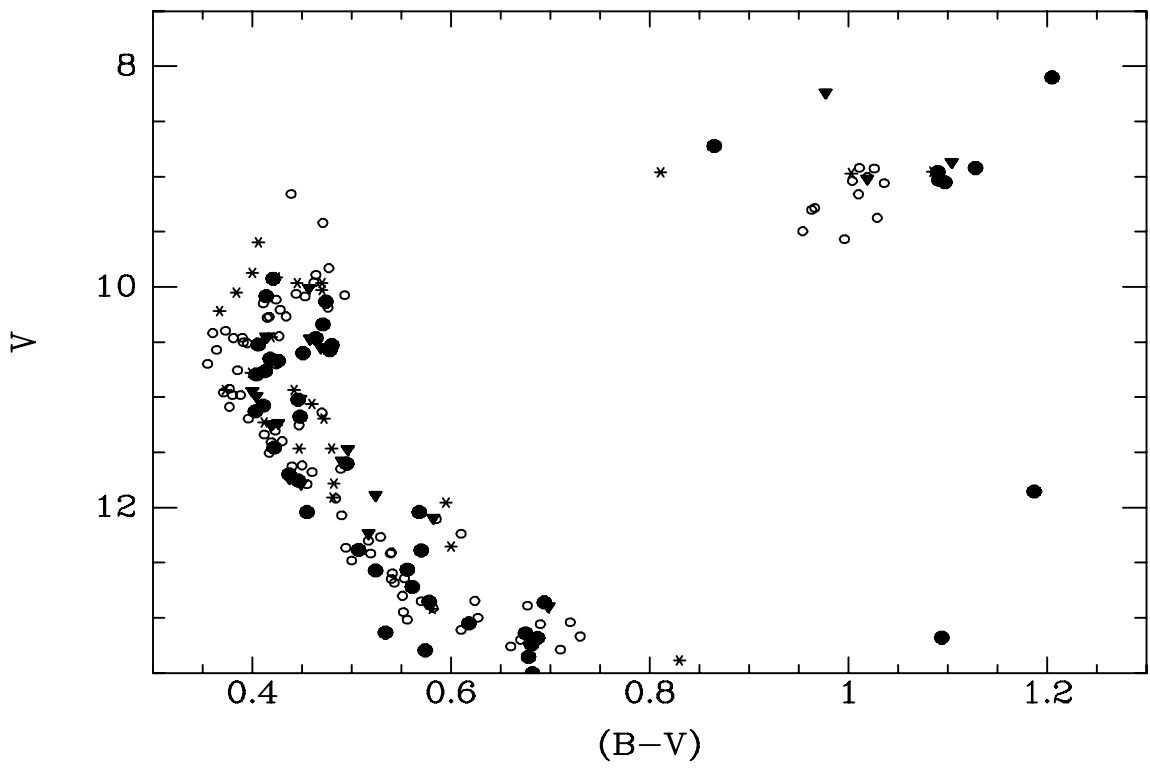
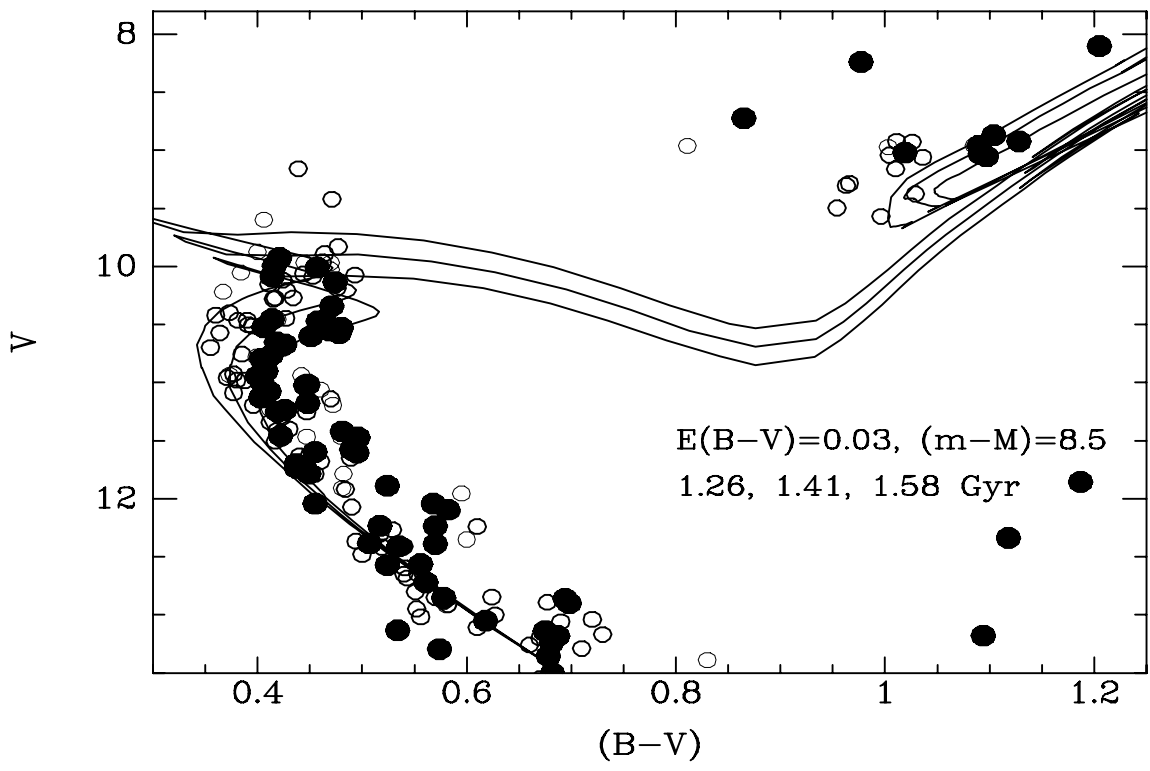
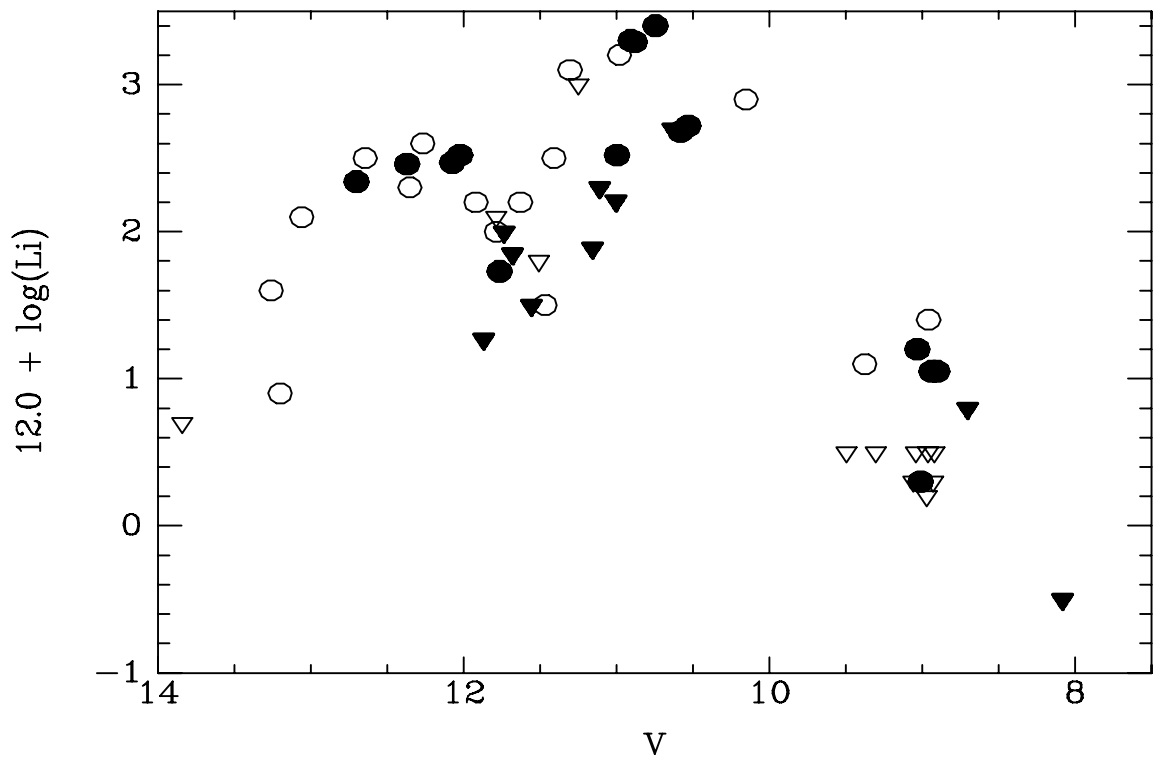
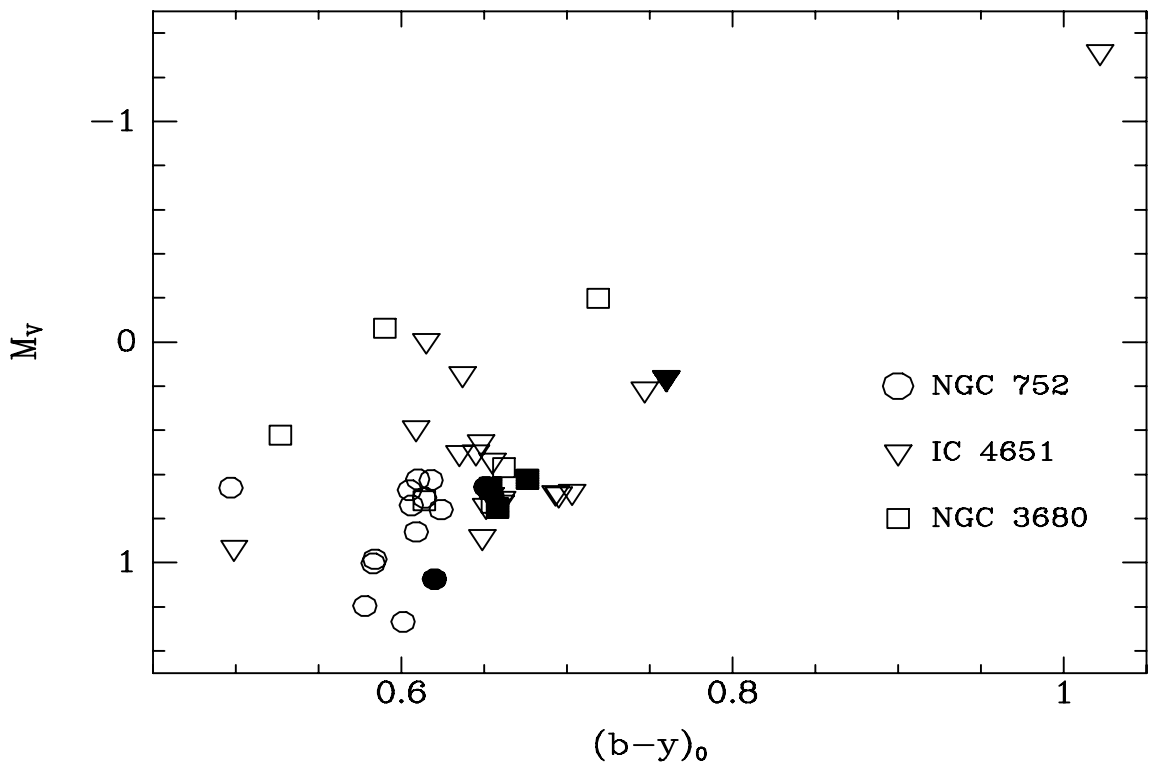
